\documentclass[aps,reprint]{revtex4-1}

\usepackage{graphicx,amsmath,bm,dcolumn}
\usepackage[utf8]{inputenc}
\usepackage[T1]{fontenc}

\newcommand{\br}{\ensuremath{\bm{r}}}
\newcommand{\vext}{\ensuremath{v_{\text{ext}}}}

\begin{document}

\title{Simple hydrogenic estimates for the exchange and correlation energies of atoms and atomic ions, with implications for density functional theory}
\author{Aaron D. Kaplan}
  \email{kaplan@temple.edu}
\affiliation{Department of Physics, Temple University, Philadelphia, PA 19122}
\author{Biswajit Santra}
\affiliation{Department of Physics, Temple University, Philadelphia, PA 19122}
\author{Puskar Bhattarai}
\affiliation{Department of Physics, Temple University, Philadelphia, PA 19122}
\author{Kamal Wagle}
\affiliation{Department of Physics, Temple University, Philadelphia, PA 19122}
\author{Shah Tanvir ur Rahman Chowdhury}
\affiliation{Department of Physics, Temple University, Philadelphia, PA 19122}
\author{Pradeep Bhetwal}
\affiliation{Department of Physics, Temple University, Philadelphia, PA 19122}
\author{Jie Yu}
\affiliation{Department of Physics, Temple University, Philadelphia, PA 19122}
\author{Hong Tang}
\affiliation{Department of Physics, Temple University, Philadelphia, PA 19122}
\author{Kieron Burke}
\affiliation{Departments of Chemistry and Physics, University of California, Irvine, CA 92697}
\author{Mel Levy}
\affiliation{Department of Chemistry and Quantum Theory Group, Tulane University, New Orleans, LA 70118}
\author{John P. Perdew}
\affiliation{Department of Physics, Temple University, Philadelphia, PA 19122}
\affiliation{Department of Chemistry, Temple University, Philadelphia, PA 19122}

\date{\today}

\begin{abstract}
  Exact density functionals for the exchange and correlation energies are approximated in practical calculations for the ground-state electronic structure of a many-electron system. An important exact constraint for the construction of approximations is to recover the correct non-relativistic large-$Z$ expansions for the corresponding energies of neutral atoms with atomic number $Z$ and electron number $N=Z$, which are correct to leading order ($-0.221 Z^{5/3}$ and $-0.021 Z \ln Z$ respectively) even in the lowest-rung or local density approximation. We find that hydrogenic densities lead to $E_{\mathrm{x}}(N,Z) \approx -0.354 N^{2/3} Z$ (as known before only for $Z \gg N \gg 1$) and $E_{\mathrm{c}} \approx -0.02 N \ln N$. These asymptotic estimates are most correct for atomic ions with large $N$ and $Z \gg N$, but we find that they are qualitatively and semi-quantitatively correct even for small $N$ and for $N \approx Z$. The large-$N$ asymptotic behavior of the energy is pre-figured in small-$N$ atoms and atomic ions, supporting the argument that widely-predictive approximate density functionals should be designed to recover the correct asymptotics. It is shown that the exact Kohn-Sham correlation energy, when calculated from the pure ground-state wavefunction, should have no contribution proportional to $Z$ in the $Z\to \infty$ limit for any fixed $N$.
\end{abstract}

\maketitle

\section{Introduction}

In this work, we will find closed-form formulas for the exchange energy and correlation energy of an atom or atomic ion with electron number $N$ and proton number $Z$. We will paint with a broad brush, seeking not the most accurate formulas but the simplest and most understandable ones, from which we can draw conclusions relevant to the construction of density functional approximations for these energies.

In exact non-relativistic quantum chemistry \cite{sza82}, the Hartree-Fock ground-state wavefunction is that single Slater determinant that minimizes the expectation value of the Hamiltonian. The quantum chemical correlation energy is the difference between the true total energy and the Hartree-Fock total energy. In exact Hohenberg-Kohn-Sham density functional theory \cite{hoh64,koh65,per78,lev79,lie83}, the Kohn-Sham ground-state wavefunction is that wavefunction that yields the true ground-state electron density and minimizes the expectation value of the kinetic energy \cite{lev85}, i.e., it is the ground eigenstate of the Kohn-Sham effective Hamiltonian. When that ground-state is degenerate, it can be a linear combination of a few Slater determinants, chosen to connect adiabatically \cite{lev85} to a given interacting ground state. When it is a single Slater determinant, the exact Kohn-Sham exchange and correlation energies of atoms and atomic ions are numerically close to those defined in quantum chemistry, and the quantum chemical correlation energy is an upper bound to the exact Kohn-Sham correlation energy \cite{gro96}. Only for one-electron densities do these two exact theories have exactly the same Slater determinant, the same exchange energy (to cancel the Hartree energy), and the same (zero) correlation energy.

In Kohn-Sham theory, the exchange and correlation energies are functionals of the electron density. Approximations to these functionals are made for the sake of practical computation for real atoms, molecules, and solids. The simplest such approximation is the local density approximation (LDA) \cite{koh65}
\begin{equation}
  E_{\nu}^{\text{LDA}}[n] = \int n(\br) \varepsilon_{\nu}^{\text{unif}}(n(\br)) d\br
\end{equation}
where $n(\br)$ is the electron density, $\nu = {\mathrm{x}}$ (exchange) or ${\mathrm{c}}$ (correlation), and $\varepsilon_{\nu}^{\text{unif}}(n)$ is the corresponding energy per electron in an electron gas of uniform density $n(\br)$. (A spin-polarized system requires $\varepsilon_{\nu}^{\text{unif}}(n_{\uparrow},n_{\downarrow})$.) Higher-rung functionals (e.g., Refs. \onlinecite{per96,sun15}) retain the correct uniform-gas limit while satisfying other exact constraints.

Although LDA is exact for a density that varies slowly over space, its relevance to real atoms and molecules is not obvious. Dirac \cite{dir30} added LDA exchange to the Thomas-Fermi model, and Schwinger \cite{sch81} may have been the first to realize that LDA exchange becomes relatively exact for neutral atoms in the limit of large atomic number. In this limit, the bulk of the density becomes Thomas-Fermi like, with a locally-slow spatial variation \cite{per06,ell08}, and the exact energies have large-$Z$ asymptotic expansions \cite{bec86,ell09,bur16,can18}
\begin{eqnarray}
  E_{\mathrm{x}}(Z,Z) &=& -A_{\mathrm{x}} Z^{5/3} + B_{\mathrm{x}} Z + ... \quad (N = Z) \label{eq:asyx} \\
  E_{\mathrm{c}}(Z,Z) &=& -A_{\mathrm{c}} Z \ln Z + B_{\mathrm{c}} Z + ... \quad (N = Z). \label{eq:asyc}
\end{eqnarray}
The leading coefficients ($A_{\mathrm{x}} = 0.221$ hartree, $A_{\mathrm{c}} = 0.021$ hartree) are known \cite{bur16,can18,kun10} to be those from the LDA evaluated on the self-consistent Kohn-Sham (or Thomas-Fermi) density for large $N = Z$, and corrections to LDA (e.g., Refs. \onlinecite{per96,sun15}) can give higher-order coefficients. Recent work \cite{san19,zop19,bha20} has shown that errors of even a few percent in the uniform-density limit can seriously undermine the accuracy of approximate density functionals for the equilibrium properties of atoms and small molecules.

There is evidence that satisfaction of Eqs. \ref{eq:asyx} and \ref{eq:asyc} can produce functionals that are notably accurate for the atomization energies of molecules, without being fitted to molecules. The Becke 1988 (B88 \cite{bec88}) generalized gradient approximation (GGA) for exchange, still widely used in chemistry, was constructed to recover LDA in the slowly-varying limit (recovering $A_{\mathrm{x}}$ of Eq. \ref{eq:asyx}) and was fitted only to the exchange energies of rare-gas atoms, but was noted to be consistent with the leading asymptotic correction to LDA exchange. Elliott and Burke \cite{ell09} showed that this functional recovers a nearly-exact (within 1.1\%) value for $B_{\mathrm{x}}$ of Eq. \ref{eq:asyx}. The Perdew-Burke-Ernzerhof (PBE \cite{per96}) GGA, while it is more accurate than B88 for solids, is less accurate for molecular atomization energies and for $B_{\mathrm{x}}$. But the acGGA \cite{can18} revision of PBE, and the SCAN meta-GGA \cite{sun15}, both constructed to satisfy Eqs. \ref{eq:asyx} and \ref{eq:asyc} (as well as other exact constraints) without fitting to molecules, are both more accurate for atomization energies than is PBE (and SCAN is remarkably more accurate).

\section{$1/Z$ perturbation expansions}

Somewhat related to the large-$Z$ expansion of Eqs. \ref{eq:asyx} and \ref{eq:asyc} is the $1/Z$ perturbation expansion \cite{mar72,tal80} of the total energy of an $N$-electron atomic ion
\begin{equation}
  E(N,Z) = Z^2\left[ \varepsilon_1(N) + \frac{\varepsilon_2(N)}{Z} + ... \right], \label{eq:ptexp}
\end{equation}
where the unperturbed problem is $N$ non-interacting electrons in the potential $-Z/r$ and the perturbation is the Coulomb repulsion among the electrons. Only a few of the coefficients in Eq. \ref{eq:ptexp} are known, and only for a few small values of $N$. In the limit $N \ll Z$, the leading terms of Eq. \ref{eq:ptexp} should be relatively accurate. Moreover, the first term exactly describes $N = 1$. In the regime $N \ll Z$, $Z$ is large enough that the electron-nucleus attraction, $-Z/r$, dominates electron-electron Coulomb repulsion, and the latter is treated as a perturbation. The unperturbed problem (i.e., the non-interacting system) has a hydrogenic density that occupies the $N$ scaled ($Z^{3/2} \psi_{i,\sigma}(Z\br)$) hydrogen-atom spin orbitals $\psi_{i,\sigma}(\br)$ of lowest energy. Clearly, for $Z \gg N-1$, the $-Z/r$ interaction of the electron with the nucleus will dominate all other terms in the Kohn-Sham one-electron potential. For $Z = N$, however, the exact Kohn-Sham potential will vary from $-Z/r$ (plus a positive constant from the Hartree and exchange-correlation potentials) to $-1/r$ as $r$ increases from 0 to $\infty$. Moreover, by the Hellmann-Feynman theorem, $Z \partial E(N,Z)/\partial Z=2Z^2\varepsilon_1(N) + Z \varepsilon_2(N) + ...$ is the electron-nuclear attraction energy.

The various components of the total energy also have $1/Z$ expansions. For example,
\begin{eqnarray}
  E_{\mathrm{x}}(N,Z) &=& \alpha_1(N) Z + \alpha_2(N) + ... \label{eq:xnz} \\
  E_{\mathrm{c}}(N,Z) &=& \beta_1(N) Z + \beta_2(N) + ... ~. \label{eq:cnz}
\end{eqnarray}
The Kohn-Sham wavefunction was discussed in the second paragraph of this article. Here the full Hamiltonian $\hat{H}$ is the sum of a hydrogenic part $\hat{H}_0$ and a weak electron-electron perturbation $\hat{V}_{\mathrm{ee}}$. In a quantum chemical calculation, $\beta_1(N)=0$ for those electron numbers $N$ (e.g., $N=1,2,3,7,8,9,10,11$) for which the ground state of $\hat{H}_0$ is either non-degenerate ($N=2,10$) or is without degenerate configurations of the same multiplet symmetry, so that the Kohn-Sham wavefunction is a single Slater determinant. These must also be electron numbers for which $\beta_1(N)=0$ in exact Kohn-Sham theory, since there is no qualitative difference between the Kohn-Sham and Hartree-Fock wavefunctions. For $N=4$, where the degenerate configurations $(1s)^2(2s)^2$ and $(1s)^2(2p)^2$ can both belong to the multiplet $^1 S$, the Kohn-Sham wavefunction is a linear combination of Slater determinants, differing significantly from the Hartree-Fock wavefunction. A different choice of Kohn-Sham representation (using an ensemble ground-state density rather than a pure ground-state wavefunction as in Appendix A) for $N=4$ and $Z \gg N$, in which the $2s$ and $2p$ orbitals are degenerate for $Z>23$, is discussed in Ref. \onlinecite{sav03}. Limiting constant values for the correlation energy can arise \cite{lev91} because the hydrogenic density $Z^3 f_N(Z r)$ is uniformly scaled to the high-density limit when $Z \to \infty$ at fixed $N$, and this uniform-scaling behavior is built into the PBE GGA \cite{per96} and the SCAN meta-GGA \cite{sun15}. We argue in Appendix A that, within the exact Kohn-Sham theory, $\beta_1(N)=0$ regardless of degeneracies. Our proof does not hold for the perturbation series of the quantum chemical correlation energy, for which it is well-known that the leading-order is $Z^0$ for a non-degenerate ground state, and $Z^1$ for some degenerate ground states \cite{per97,hol11}. The exact Hartree-Fock and exact Kohn-Sham exchange energies have different uniform scaling behaviors \cite{lev85}, and their wavefunctions differ substantially in the case of a degenerate ground-state. Thus we can expect different large-$Z$ limits of their corresponding correlation energies.

The quantum chemical coefficients $\alpha_1(N)$ and $\beta_2(N)$ for small $N$ are reported in Tables I and II of Ref. \onlinecite{sta04}. Our Table \ref{tab:ascf} shows that $\alpha_1(N)/N^{2/3}$ and $\beta_2(N)/(N \ln N)$ are nearly independent of $N$ (not noticed in Ref. \onlinecite{sta04}), suggesting that $\alpha_1(N) \sim N^{2/3}$, and $\beta_2(N) \sim(N \ln N)$. Here, we use $\sim$ to denote the leading-order behavior of a function. Thus, at least for $N \ll Z$ and for $N$ such that $\beta_1(N) = 0$,
\begin{eqnarray}
  E_{\mathrm{x}}(N,Z) &\approx& -0.354 N^{2/3} Z \label{eq:exest} \\
  E_{\mathrm{c}}(N,Z) &\approx& -0.02 N \ln N. \label{eq:ecest}
\end{eqnarray}
The value $-0.354$ is the analytic large-$Z$ limit of $\alpha_1(N)/N^{2/3}$, as explained around Eq. \ref{eq:lhasy2}. The value $-0.02$ is a roundoff of all the values of $\beta_2(N)/(N\ln N)$ for $N>2$ from Table \ref{tab:ascf}, accounting for the larger uncertainty in the numeric values of this coefficient. Now, setting $N = Z$ in Eqs. \ref{eq:exest} and \ref{eq:ecest} leads to a result that is qualitatively and semi-quantitatively like the leading terms of Eqs. \ref{eq:asyx} and \ref{eq:asyc}. In particular, $E_{\mathrm{x}}(Z,Z) \sim Z^{5/3}$ and $E_{\mathrm{c}}(Z,Z) \sim Z \ln Z$. The primary difference is in the coefficient of $Z^{5/3}$, in part because the density functional for the exchange energy depends on the detailed shape of the electron density (e.g. hydrogenic vs. self-consistent neutral Thomas-Fermi) but not so for correlation in leading order. Appendix B presents a simple derivation of the leading-order terms in the asymptotic series for the correlation energy, and shows that they are identical for hydrogenic and self-consistent neutral Thomas-Fermi densities.

The more important conclusion, which will be validated and applied in the rest of this article, is that the large-$N$ asymptotics of the exchange and correlation energies for $Z \gg N$ are discernible even for small $N$. A corollary to this is that the large-$N$ asymptotics can be roughly estimated from the small-$N$ energetics.

\begin{table}
  \caption{The leading coefficients $\alpha_1(N)$ and $\beta_2(N)$ (both in hartree) in Eqs. \ref{eq:xnz} and \ref{eq:cnz}, from Ref. \onlinecite{sta04}, divided by the displayed functions of electron number $N$. \label{tab:ascf}}
  \begin{ruledtabular}
    \centering
    \begin{tabular}{ccc}
      N & $\alpha_1(N)/N^{2/3}$ & $\beta_2(N)/(N \ln N)$ \\ \hline
      2 & -0.3937 & -0.0337 \\
      3 & -0.3471 & -0.0163 \\
      7 & -0.3508 & -0.0174 \\
      8 & -0.3569 & -0.0184 \\
      9 & -0.3658 & -0.0187 \\
      10 & -0.3766 & -0.0186 \\
      11 & -0.3647 & -0.0173 \\
    \end{tabular}
  \end{ruledtabular}
\end{table}

Conlon \cite{con83} showed that, in the limit $Z = N \to \infty$, the Hartree-Fock exchange energy for any Coulomb system tends to the LDA exchange energy evaluated on the self-consistent Thomas-Fermi density. Thus, there is no inherent contradiction in the values of the exchange coefficient: hydrogenic densities are not the correct $Z \approx N \to \infty$ limits. The Thomas-Fermi approximation to the hydrogenic density, $n_{\mathrm{h}}^{\text{TF}}$, is known analytically (refer to Eq. \ref{eq:tfhd} in Appendix B or Refs. \onlinecite{mar76,eng88,bur16}). Evaluating the LDA on the Thomas-Fermi approximation for the hydrogenic density of a neutral atom yields, for $N=Z$,
\begin{equation}
  E_{\mathrm{x}}^{\text{LDA}}[n_{\mathrm{h}}^{\text{TF}}] = \left(\frac{2}{3} \right)^{1/3} \frac{4}{\pi^2} Z^{5/3} \approx -0.354 Z^{5/3}. \label{eq:lhasy}
\end{equation}
For heavy positive ions, where $N < Z \to \infty$, the LDA exchange energy evaluated on the hydrogenic Thomas-Fermi density tends to \cite{dmi75,mar76}, for $N \leq Z$,
\begin{equation}
  E_{\mathrm{x}}^{\text{LDA}}[n_{\mathrm{h}}^{\text{TF}}] = -0.354 N^{2/3} Z \left[1 + \mathcal{O}\left(\frac{N}{Z}\right)\right]. \label{eq:lhasy2}
\end{equation}
A numeric estimate of this coefficient from Ref. \onlinecite{sny09} on a neutral hydrogenic density of 2030 electrons agreed precisely with the analytic values of Eqs. \ref{eq:lhasy} and \ref{eq:lhasy2}.

The exact exchange energy for a given spin-unpolarized density $n(\br)$ is expected to be bounded by the conjectured tight lower bound \cite{per14,sun15a}
\begin{equation}
  E_{\mathrm{x}}[n] \geq 1.174 E_{\mathrm{x}}^{\text{LDA}}[n],
\end{equation}
satisfied for all spin-unpolarized densities by LDA and SCAN \cite{sun15}, but not by PBE \cite{per96}. This bound holds rigorously for a spin-unpolarized two-electron density \cite{per14}. LDA is expected to be relatively less accurate for such densities than for spin-unpolarized densities with $N>2$, and no violation of the conjectured bound is known. LDA typically becomes relatively exact as more electrons are packed into a given volume of space. We give an alternative derivation of the exactness of LDA in the large fixed $N$ and $Z \to \infty$ limit in Appendix B, based on a scaling argument.

Eq. \ref{eq:xnz} suggests that the leading correction $B_{\mathrm{x}} Z$ of Eq. \ref{eq:asyx} arises in part from the two $1s$ electrons that are present in any atom of large $Z$ and for which the density never becomes slowly-varying. This is analogous to the Scott correction \cite{sco52} to Thomas-Fermi theory. Within Thomas-Fermi theory for atoms and ions, the majority of electrons are within a distance $Z^{-1/3}$ from the nucleus, whereas the Scott correction applies \cite{hei95} to electrons within a distance $Z^{-1}$. In the limit $N=Z \to \infty$ and at distances short compared to the Thomas-Fermi scale $Z^{-1/3}$, the density is approximately independent of $N$, a Bohr-like atom \cite{hei95}.

In the limit $1 \ll N \ll Z$, the Thomas-Fermi hydrogenic density becomes slowly-varying on the scale of the exchange energy, but not on the scale of the correlation energy. This is demonstrated in Appendix B using scaling arguments. The LDA is by definition exact for any uniform density, thus the LDA exchange energy is exact as $Z \to \infty$ (where the density becomes high and locally uniform). This argument cannot determine the exactness of the LDA correlation energy, but suggests that the convergence of the exact correlation energy to the LDA in the limit $1 \ll N \ll Z$ is slower than for the exchange energy. It is known \cite{kun10} from a direct semiclassical calculation that the LDA correlation energy is exact to leading-order for heavy neutral atoms $N = Z \to \infty$. At next-to-leading order, the LDA coefficient is of the wrong sign \cite{bur16}, motivating the need for gradient corrections for real systems \cite{ma68}.

\section{Conclusions}

\begin{figure}
  \centering
  \includegraphics[width=\columnwidth]{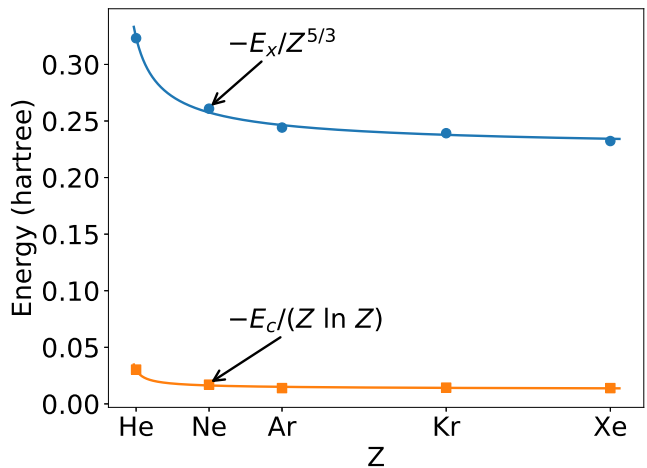}
  \caption{The scaled exact exchange \cite{bec88} (blue) and scaled exact correlation energies \cite{bur16} (orange) of the non-relativistic, neutral rare gas series. The smooth curves are a guide for the eye, and not intended to be extrapolative. \label{fig:rgas}}
\end{figure}

Figure \ref{fig:rgas} shows that the {\it exact} exchange energies \cite{bec88} of the neutral rare-gas atoms divided by $Z^{5/3}$ vary from an almost-hydrogenic value at $Z = 2$ to a more Thomas-Fermi-like value at $Z = 54$. The more compact hydrogenic density has a more negative exchange energy for a given $Z = N$. The {\it exact} correlation energies \cite{bur16} divided by $Z \ln Z$ in Fig. \ref{fig:rgas} show an almost constant value of -0.014 from $Z$ = 18 to 54. Interestingly, a much better coefficient of $Z\ln Z$ for neutral atoms of large $Z$ can be found by approaching from the fixed $N$, large $Z$ direction, as shown in Table \ref{tab:ascf} and Appendix B. For the series of neutral atoms, one needs to extrapolate carefully \cite{bur16,can18,san19} to much larger $Z$ to approach the asymptotic limit, but that limit is clearly pre-figured in the energies of real atoms. This suggests that widely-predictive approximate functionals should be constrained to recover the correct large-$Z$ asymptotics. And, in fact, LDA \cite{koh65}, PBE \cite{per96}, SCAN \cite{sun15}, and acPBE \cite{can18} sequentially improve \cite{sun15,can18} the large-$Z$ asymptotics. Functionals that satisfy sufficient exact constraints, including but not limited to Eqs. \ref{eq:asyx} and \ref{eq:asyc} with correct coefficients, can be predictive over the wide space of atoms, molecules, and solids, without being fitted to any bonded system. While that goal is not yet reached, the successes of SCAN \cite{sun16,che17,goe17,fur18,sha18,zha18} suggest that it can be.

Moreover, in the non-relativistic $Z \to \infty$ limit, the periodic table becomes perfectly periodic, with limiting first ionization energies for each column that increase across each row and for which the local density approximation for exchange and correlation may become relatively exact \cite{con10}.

The explanations given here for the fundamental relevance of the LDA and its generalizations to real atoms and molecules were pioneered in work \cite{per06,ell08,bur16,can18} which focuses on LDA's correctness for neutral atoms, molecules, and solids of large atomic number. They constitute a third wave of such explanations. The first wave focused on LDA's satisfaction of exact constraints on the exchange-correlation hole around an electron \cite{lan75,gun76,bur98}. The second wave, following Refs. \onlinecite{per96,sun15}, focused on the fact that LDA inherits many (about nine) exact constraints on the density functional for the exchange-correlation energy from its appropriate norm, the uniform electron gas.

Our observation, that the large-$N$ asymptotics of the exchange-correlation energy are ``writ small'' in the energies of atoms and atomic ions of very small $N$, is a contribution in support of this third wave of explanations. This prefiguration in small-$Z$ neutral atoms is known \cite{bec86} but not always emphasized \cite{bec88,ell09,jon15}. (The reasonable accuracy of LDA exchange for neutral atoms of small $Z$ is also shown, for example, in Fig. 3 of Ref. \onlinecite{jon15}.) The large-$Z$ asymptotes by themselves are insufficient, since a pseudopotential method can remove them without significantly changing the energetics of the valence electrons, but Ref. \onlinecite{con10} demonstrated its importance for energy differences.

Recent work has shown that the Perdew-Zunger self-interaction correction (PZ SIC), which makes any approximate density functional exact for non-overlapped one-electron densities, introduces errors of as much as 6\% in the large-$Z$ or slowly-varying-density limit \cite{san19}, and that these errors can degrade the predicted equilibrium properties of molecules. Carefully scaling down the PZ SIC correction to zero in slowly-varying regions yields much better results \cite{zop19,bha20}. The LYP correlation functional is highly accurate for first and second row molecules, but highly inaccurate for the uniform gas. To be appropriate for both solids and molecules, a density functional for exchange and correlation should recover LDA in the uniform limit, and correct LDA for finite systems.  Beyond that, Ref. \onlinecite{can18} demonstrated that small modifications of the PBE functional (which in its original form recovers LDA exactly in the appropriate limit, and roughly predicts the next-order terms of the large-$Z$ expansion), to recover exactly the next-order terms in Eqs. \ref{eq:asyx} and \ref{eq:asyc}, moderately improve the PBE atomization energies of molecules. In future work, we hope to test the idea that recovery of the next-order terms in Eqs. \ref{eq:asyx} and \ref{eq:asyc} can be used more generally to improve approximate density functionals.

\begin{acknowledgments}
The work of ADK was supported by the Department of Energy (DOE), Office of Science (OS), Basic Energy Sciences (BES), through grant no. DE-SC0012575 to the Energy Frontier Research Center: Center for Complex Materials from First Principles. The work of BS was supported by DOE, OS, BES under grant no. DE-SC0018331. The work of PB, KW, and JPP was supported by the U.S. National Science Foundation under grant no. DMR-1939528. The work of TURC and HT was supported by DOE, OS, BES under grant no. DE-SC0018194. KB was supported by DOE under grant no. DE-FG02-08ER46496. KB thanks John Snyder, Jeremy Ovadia, Krishanu Ray, and Timothy Middlemas, whose unpublished notes contributed to Appendix \ref{sec:app_b}. We thank a referee for suggesting the possibility that $\beta_1(N)=0$ in Eq. \ref{eq:cnz} for all $N$ within Kohn-Sham theory.
\end{acknowledgments}

\section*{Data Availability}

The data that supports the findings of this study are available within the article.

\appendix

\section{The $1/Z$ expansion of the exact Kohn-Sham correlation energy starts at order $Z^0$\label{sec:app_a}}

Consider the Hamiltonian $\hat{H} = \hat{H}_0 + \hat{V}_{\mathrm{ee}}$, where $\hat{H}_0 = \hat{T} + \hat{V}_{\text{ext}}$. Here, $\hat{T}$ is the kinetic energy operator, $\hat{V}_{\text{ext}}$ is the external potential or Coulomb attraction to the nuclei, and $\hat{V}_{\mathrm{ee}}$ is the electron-electron Coulomb repulsion.

The exact Kohn-Sham correlation energy is defined as \cite{lev85}
\begin{equation}
  E_{\mathrm{c}} = \langle \Psi_n|( \hat{T} + \hat{V}_{\mathrm{ee}})| \Psi_n \rangle - \langle \Phi_n|( \hat{T} + \hat{V}_{\mathrm{ee}})| \Phi_n \rangle.
\end{equation}
Here $\Psi_n$, the ground-state wavefunction, minimizes the expectation value of $\hat{H}$ and defines the ground-state density $n(\br)$, while $\Phi_n$, the exact Kohn-Sham wavefunction, is that ground eigenstate of the non-interacting or Kohn-Sham Hamiltonian $\hat{H}_{\text{KS}}$ (a linear combination of at most a few Slater determinants) that yields the same ground-state density $n(\br)$. Consequently,
\begin{equation}
  E_{\mathrm{c}} = \langle \Psi_n|\hat{H}| \Psi_n \rangle - \langle \Phi_n|\hat{H}| \Phi_n \rangle. \label{eq:ec_exact}
\end{equation}
Now write $\Phi_n=  \Psi_n+ \delta \Psi$, and expand everything (including the ground state density $n(\br)$) in powers of $\hat{V}_{\mathrm{ee}}$ via degenerate perturbation theory. As $\hat{V}_{\mathrm{ee}} \to 0$, $\Phi_n$ tends to the right linear combination of degenerate ground states of $\hat{H}_0$, and $\delta\Psi$ tends to zero like $\hat{V}_{\mathrm{ee}}$.  Since $\Psi_n$ minimizes the expectation value of $\hat{H}$, the leading term of the correlation energy is of order $(\delta \Psi)^2$ or $\hat{V}_{\mathrm{ee}}^2$. In the $1/Z$ expansion of Eq. \ref{eq:ptexp}, this is a contribution of order $Z^2(1/Z)^2 = Z^0$.

Note that the exact Kohn-Sham exchange energy \cite{lev85} $\langle \Phi_n|\hat{V}_{\mathrm{ee}}| \Phi_n \rangle - U_{\mathrm{H}}[n]$, where $U_{\mathrm{H}}[n]$ is the Hartree electrostatic energy, can also differ substantially from its quantum chemical counterpart, the Hartree-Fock exchange energy.

Therefore, in the perturbation series for the total energy, the term linear in $Z$ (known precisely from quantum chemical calculations \cite{per97}) must appear in other components of the Kohn-Sham expansion (e.g., the exact Kohn-Sham exchange energy or exact Hartree potential).

In the standard degenerate or non-degenerate perturbation expansion of Eqs. \ref{eq:ptexp}-\ref{eq:cnz}, the density changes along with the coupling constant $\alpha$ that multiplies $\hat{V}_{\mathrm{ee}}$. A different perturbation expansion, in which the density is fixed at its physical or $\alpha=1$ value, was proposed by G\"{o}rling and Levy (GL). In non-degenerate GL perturbation theory, the coefficient of the leading or $\alpha^2$ contribution to the correlation energy is given by Eq. 4 of Ref. \onlinecite{gor93}.

~\\

\section{Hydrogenic Ions with $1 \ll N \ll Z$\label{sec:app_b}}

In the heavy ion limit, with $N$ fixed to a large value and $Z \to \infty$, noninteracting Thomas-Fermi theory becomes relatively exact \cite{tal80,eng88}. One can show that the density of non-interacting Thomas-Fermi theory is given by
\begin{equation}
  n(\br) = \frac{1}{3\pi^2}\left[2(\mu -  \vext(\br)) \right]^{3/2} \Theta(\mu -  \vext(\br))
\end{equation}
where the chemical potential $\mu$ is a Lagrange multiplier determined by $\int n(\br) d^3 r = N$, and $\Theta$ is a step function defining the turning surface. Let
\begin{equation}
  \nu = N/Z
\end{equation}
be the degree of ionization such that $0 < \nu \leq 1$. We do not consider the case $N > Z$, as it has been found that $N \leq Z + 1$ for real atoms \cite{ben85}. In the spherical case, one can rewrite the non-interacting Thomas-Fermi density in terms of the dimensionless variable $x = r/r_c$, with $r_c$ the turning surface radius. For hydrogenic densities, $\vext(r) = -Z/r$, and thus $r_c = -Z/\mu$ (NB: $\mu < 0$ in this case as $\vext(r) < 0$ for all $r$). Constraining the density to integrate to $N$ determines $r_c = (18 N^2)^{1/3}/Z$, allowing the density to be rewritten as \cite{hei95,bur16}
\begin{equation}
  n(x) = \frac{Z^2}{\nu}\frac{2}{9\pi^2}(1/x - 1)^{3/2} \Theta(1-x). \label{eq:tfhd}
\end{equation}
In the limit $Z \gg N - 1$, a hydrogenic density built up from $Z$-scaled hydrogen-atom orbitals obeying the hydrogen-atom aufbau principle becomes relatively exact. With the additional condition $N \gg 1$, Eq. \ref{eq:tfhd} becomes relatively exact almost everywhere, so it usefully imitates a hydrogenic density with $N \gg 1$. Here, ``almost everywhere'' means a region excluding electrons in the density tail or very near the nucleus, but including all of the electrons in the order of limits in which $Z \to \infty$ is followed by $N \to \infty$. As we lack a simple, closed-form expression \cite{hei95} for a hydrogenic density constructed from hydrogen-atom orbitals in the large $N$ limit, the non-interacting Thomas-Fermi density is needed to derive the asymptotic properties of hydrogenic densities.

From the scaling of $n(x)$ with $Z$ and $\nu$, we see that the Wigner-Seitz radius $r_s(\br) = [4\pi n(\br)/3]^{-1/3}$ scales like
\begin{equation}
  r_s(x) = \frac{\nu^{1/3}}{Z^{2/3}} \frac{3\pi^{1/3}}{2} (1/x - 1)^{-1/2}  \Theta(1-x).
\end{equation}
In the heavy ion limit, as $N < Z$ with $Z \to \infty$, $r_s$ tends to zero (the ``high-density limit'') except near $x=1$. In the heavy neutral atom limit, $N = Z \to \infty$, $r_s \to 0$ as well. Note also that $r_c \to 0$ as $Z \to \infty$, implying that the density localizes near the origin in this limit.

The LDA is exact for any uniform electron density, and is relatively exact for any slowly-varying electron density. We say that a density is slowly-varying when its dimensionless gradients are less than order one. For exchange, the appropriate dimensionless density-gradient is $s(\br)$, defined on the scale of the Fermi wavevector $k_F(\br) = [3\pi^2 n(\br)]^{1/3}$,
\begin{equation}
  s(\br)  = \frac{|\nabla n(\br)|}{2(3\pi^2)^{1/3} n(\br)^{4/3}},
\end{equation}
and for correlation, the appropriate density-gradient is $t(\br)$, defined on the scale of the Thomas-Fermi screening wavevector $k_s = \sqrt{4 k_F/\pi}$,
\begin{equation}
  t(\br)  = \left(\frac{3\pi^2}{16}\right)^{1/3} \frac{\phi(\zeta) s(\br)}{r_s(\br)^{1/2}}.
\end{equation}
Here, $\phi(\zeta)$ is a function of the spin-polarization $\zeta$ defined in Ref. \onlinecite{per96}. We know that the density is effectively spin-unpolarized in the limit of large $N$, for which $\phi(\zeta = 0) = 1$ and can be ignored throughout. The LDA exchange (correlation) energy will be relatively exact in the heavy atom limit if $0 < s(\br) \ll 1$ ($0 < t(\br) \ll 1$), and will be exact if $s(\br) \to 0$ ($t(\br) \to 0$) as $Z \to \infty$. These are presumptively sufficient conditions to determine the exactness of the LDA; the LDA may still be relatively exact even if they are violated.

From Eq. \ref{eq:tfhd},
\begin{widetext}
\begin{equation}
  |\nabla n(x)| = \frac{Z^{7/3}}{ \nu^{5/3}} \frac{4}{18^{4/3}\pi^2}\left[ \frac{3}{2 x^2}(1/x - 1)^{1/2} \Theta(1-x) + (1/x - 1)^{3/2} \delta(1-x) \right];
\end{equation}
\end{widetext}
the delta function is irrelevant, as $x$ approaches 1 from below, thus the dimensionless gradients scale as
\begin{widetext}
\begin{eqnarray}
  s(x) &=&  Z^{-1/3} \nu^{-1/3} ~ \left( \frac{3}{16} \right)^{2/3} x^{-2}(1/x - 1)^{-3/2} \Theta(1-x) \\
  t(x) &=& \nu^{-1/2}  ~ \frac{1}{8} \left( \frac{3 \pi}{2} \right)^{1/2} x^{-2}(1/x - 1)^{-5/4} \Theta(1-x).
\end{eqnarray}
\end{widetext}
Both $s$ and $t$ are divergent as $x \to 0$ and $x \to 1$. However, for $0 < x < 1$, $s(x) \to 0$ as $Z \to \infty$ in both the heavy ion and neutral atom limits, thus the LDA exchange energy becomes relatively exact as $Z \to \infty$. This constitutes a simple argument for the exactness of the LDA exchange energy as found by Refs. \onlinecite{dmi75,con83}. But, in the neutral atom limit, $t(x) \sim \mathcal{O}(1)$, not characteristic of a slowly-varying density. As the condition $ 0 \leq t \ll 1$ is only sufficient, our scaling argument cannot determine if the LDA correlation energy is exact in the heavy neutral atom and heavy ion limits.

This scaling analysis, applied to the self-consistent Thomas-Fermi density of a neutral atom, shows that $s \sim Z^{-1/3}$, but $t \sim \mathcal{O}(1)$ as $Z \to \infty$. As noted in Ref. \onlinecite{per06}, LDA still seems to get the correct leading term in Eq. \ref{eq:asyc}, although PBE (which employs $t$) preserves the correct leading term and improves the next term in Eq. \ref{eq:asyc}. Ref. \onlinecite{per06} also shows that $s$ is already rather small in the closed-shell atoms Kr and Rn.

The coefficient of the leading-order terms in Eq. \ref{eq:asyc} can be determined by the scaling properties of a given density. Moreover, the scaling behavior of the density is sufficient to determine if the LDA exchange and correlation energies are separately exact in the high-density limit. In the high-density spin-unpolarized limit, the LDA correlation energy density tends to $\varepsilon_c^{\text{LDA}} = c_0 \ln r_s - c_1$, where $c_0 = (1 -\ln 2)/\pi^2 \approx 0.0310907$ and $c_1 \approx 0.046644$ are known from many-body perturbation theory \cite{gel57}. Consider a family of densities $n(r/r_c)$ that can be written as a function of the dimensionless position variable $r/r_c$, with $r_c$ the turning-surface radius defined by $\mu = \vext(r_c)$. The density may be expressed in powers of the nuclear charge $Z$, the ionization degree $0 < \nu = N/Z \leq 1$, and a position-dependent function $f(r/r_c)$,
\begin{equation}
  n(r/r_c) = Z^{\alpha} \nu^{\beta} f(r/r_c)
\end{equation}
with $\alpha \geq 0$ and $\beta$ real numbers. The high-density expansion of the LDA correlation energy evaluated on this family of densities yields an asymptotic series of the form
\begin{equation}
  E_{\mathrm{c}}^{\text{LDA}} =  -\frac{c_0}{3} N \left[ \alpha \ln Z  + \beta \ln \nu \right] - K^{\text{LDA}} N. \label{eq:clda}
\end{equation}
$K^{\text{LDA}}$ is a constant dependent upon $f(r/r_c)$. For a self-consistent Thomas-Fermi density ($\alpha = 2$ and $\beta = 0$) in the neutral limit ($\nu = 1$), this series yields the coefficient $A_{\mathrm{c}} = -2c_0/3 = -0.020727$ in Eq. \ref{eq:asyc}. It was shown in Eq. \ref{eq:tfhd} that the non-interacting Thomas-Fermi hydrogenic density satisfies this scaling behavior with $\alpha=2$ and $\beta = -1$, thus its leading order asymptotic series of Eq. \ref{eq:asyc} is identical to that of the self-consistent Thomas-Fermi density in the neutral limit.

A concern might come to mind: Eq. \ref{eq:clda} has explicit $Z$ dependence, but the perturbation series of Eq. \ref{eq:cnz} suggests that the exact correlation energy should not depend upon $Z$, when there are no emerging degeneracies. Recasting Eq. \ref{eq:clda} as
\begin{equation}
  E_{\mathrm{c}}^{\text{LDA}} = -\frac{c_0 \alpha}{3} N \ln N - \left[ \frac{c_0}{3} (\beta-\alpha) \ln \nu + K^{\text{LDA}} \right]N \label{eq:cldan}
\end{equation}
allows for a more direct comparison with Eq. \ref{eq:cnz}. The dominant or $N \ln N$ term in Eq. \ref{eq:cldan} is consistent with the numeric results of Table \ref{tab:ascf}. The next or $N$ term is known to be incorrect even for the neutral case. The LDA correction applies at best only to the limit $Z \to \infty$ with $N/Z$ fixed, and not to the limit $Z \to \infty$ with $N$ fixed. There exists an exact $K$ that is not recovered accurately by the LDA.

Note that the scaling behavior shared by hydrogenic and self-consistent Thomas-Fermi densities, both derived from potentials that depend linearly upon $Z$, is not universal. For an isotropic harmonic oscillator potential $\vext(r) = k r^2/2$, where $k > 0$, the non-interacting Thomas-Fermi density is \cite{bra01}
\begin{widetext}
\begin{equation}
  n(r/r_c) = k^{3/4} N^{1/2} \frac{24^{1/2}}{3\pi^2}[1 - (r/r_c)^2]^{3/2}\Theta(1 - r/r_c), \qquad r_c = \left(\frac{24 N}{k^{3/2}} \right)^{1/6}
\end{equation}
\end{widetext}
which yields a coefficient $-5c_0/12$ of the $N \ln N$ term of Eq. \ref{eq:cldan}, when $k=Z$ in the ``neutral'' limit ($\nu = 1$). As a harmonic oscillator potential with linear dependence on $Z$ is a mathematical artifice, no physical interpretation can be ascribed to the neutral limit. Setting $k = Z^4$ would recover a $1/Z$ expansion for the total energy of the same form as Eq. \ref{eq:ptexp}.


%

\end{document}